\documentclass[aps,preprint,epsfig,rotate]{revtex4}
\usepackage{tikz}
\usepackage{verbatim}
\usetikzlibrary{calc,arrows,decorations.pathmorphing,intersections}
\usepackage[font={small,sf},labelfont={bf},labelsep=endash]{caption}
\usepackage{sansmath}

\begin{document}

\title{Bound state spectrum of the triplet states in the Be atom}

\author{Alexei M. Frolov}
\email[E--mail address: ]{afrolov@uwo.ca}

\affiliation{Department of Applied Mathematics \\
 University of Western Ontario, London, Ontario N6H 5B7, Canada}

\author{Mar\'{\i}a Bel\'en Ruiz}
\email[E--mail address: ]{maria.belen.ruiz@fau.de}

\affiliation{Department of Theoretical Chemistry, \\
Friedrich-Alexander-University Erlangen-N\"urnberg, Egerlandstra\ss e 3, 
D-91058, Erlangen, Germany}

\date{\today}

\begin{abstract}

The bound state spectrum of the low-lying triplet states in the Be atom is investigated.
In particular, we perform accurate computations of the bound triplet $S$, $P$, $D$, $F$, $G$, 
$H$ and $I$ states in the Be atom. The results of these calculations are employed to 
draw the spectral diagram which contains the energy levels of the triplet states. 
Based on our computational results we can observe transitions from the low-lying bound states 
to the weakly-bound Rydberg states. For the $2^3S$, $3^3S$ and $4^3S$ states in the Be atom 
we also determine a number of bound states properties.
 
\end{abstract}

\maketitle
\newpage

\section{Introduction}

In this study we investigate the bound state spectrum of the low-lying triplet states in the four-electron Be 
atom. As is well known (see, e.g., \cite{Sob}) the bound state spectra of the four-electron Be atom and Be-like
ions include the two separate series: singlet states and triplet states. In the lowest order approximation of the 
fine-structure constant $\alpha$, where $\alpha = \frac{e^2}{\hbar c} \approx \frac{1}{137}$, there are no optical 
transitions between these two series of bound states. The bound triplet states in different four-electron 
atoms and ions have been neglected for quite some time, while various theoretical, computational and experimental 
works performed recently for the four-electron atoms and ions were mainly oriented towards the singlet states in such 
systems. In contrast with the singlet states in the four-electron atoms and ions, for the triplet bound states only a 
few highly accurate results are now known (see, e.g., \cite{Fro1}, \cite{Galv} and references therein). 

On the other hand, different triplet bound states in the four-electron atoms and ions are of great interest in a large 
number of physical problems, including spectral analysis of the stellar and laboratory plasmas, accurate prediction of the 
properties of the light element plasmas at high temperatures and arbitrary pressures, etc. For instance, many spectral lines 
which correspond to the four-electron B$^{+}$, C$^{2+}$, N$^{3+}$ and O$^{4+}$ ions are observed in the emission spetra of 
the hot Wolf-Rayet stars (see, e.g., \cite{Beals}, \cite{Allen1} and references therein). The Wolf-Rayet stars are known as 
the brightest radiating objects in our Galaxy \cite{Sobolev}. They can be used in the future to create a very accurate and
reliable navigation system in our Galaxy. 

Currently, the bound state spectrum of the triplet states in the four-electron atomic systems (atoms and
ions) is not a well known subject, and the present work provides important data in this area. The total energies of many excited states
are known only approximately. The order of different bound (triplet) states is wrongly predicted, and only approximately known. 
Published spectroscopic papers (see, e.g., \cite{Kramida}) do not include any of the rotationally excited bound states with $L 
\ge 4$. This situation has motivated us to perform accurate numerical computations of various $S(L = 0)$, $P(L = 
1)$, $D(L = 2)$, $F(L = 3)$, $G(L = 4)$, $H(L = 5)$ and $I(L = 6)$ states in the Be atom with the infinitely heavy nucleus. Such a system is usually 
designated as the ${}^{\infty}$Be atom. As follows from Table 4.1 given in \cite{FrFish} the bound states in various four-electron 
atoms and ions are difficult atomic systems for the Hartree-Fock method and all methods based on the Hatree-Fock approximation. Results 
of our calculations for excited states are significantly more accurate than analogous results known from Hartree-Fock based methods.

Based on the computed total energies of different bound (triplet) states we were able to draw the diagram of the triplet states 
bound spectrum of the four-electron Be atom (see, Fig.1). In general, to draw such a spectral diagram one needs to use accurate 
computational results for a large number of bound states. Such a spectral diagram was never presented in earlier papers which usually 
deal with one, or very few bound states in the beryllium atom. In contrast with these methods our approach allows one to determine and 
investigate all low-lying triplet states in the Be atom. In addition, with the computational results of this paper one can observe the actual 
transition from the low-lying bound states to the weakly-bound (or true Rydberg) states in the triplet series of the bound state 
spectrum of the Be atom. For the $2^3S$, $3^3S$ and $4^3S$ states in the ${}^{\infty}$Be atom 
we also perform a separate series of highly accurate computations. The results of such calculations are used to predict a number of 
bound state properties in these two states.  Concluding remarks can be found in the final Section.

\section{Hamiltonian and bound state wave functions in CI-method}  

The Hamiltonian $H$ of the four-electron atomic system is written in the form
\begin{eqnarray}
 H = -\frac{\hbar^2}{2 m_e} \Bigl[ \sum^{4}_{i=1} \nabla^2_i + \frac{1}{M_n} 
 \nabla^{2}_{5} \Bigr]
 - \sum^4_{i=1} \frac{Q e^2}{r_{i5}} +  \sum^{3}_{i=1} \sum^{4}_{j=2 (j>i)} 
 \frac{e^2}{r_{ij}} \label{eq1}
\end{eqnarray}
where $\hbar$ is the reduced Planck constant, $m_e$ is the electron mass, $e$ is the absolute value of the electron 
charge of the electron. Also, in this equation $Q$ and $M_n$ are the electric charge and mass of the nucleus expressed in 
$e$ and $m_e$, respectively. It is clear that $M_n \gg 1$ and some bound states in the four-electron atoms/ions can be 
observed, if $Q > 3$ in Eq.(\ref{eq1}). This means that the ground (singlet) state in the Li$^{-}$ ion is bounded, while
the negatively charged He$^{2-}$ ion does not exist as a bound system in any state. 

In Eq.(\ref{eq1}) and everywhere below the subscript 5 denotes the atomic nucleus, while 
subscripts 1, 2, 3 and 4 stand for electrons. As mentioned above the four-electron Be atom has two independent series of 
bound states: singlet states and triplet states. The ground singlet state with $L = 0$ (also called the $2^1S$ state) has 
the lowest total energy $E$ = -14.667 355 3(5) $a.u.$ The total energies of the two bound $2^1P(L = 1)$ and $2^3P(L = 
1)$ states are: $E$ = -14.473 44 $a.u.$ and $E$ = -14.567 24 $a.u.$ (see discussion and references in \cite{Galv,Chung}), respectively. 
It follows from these energy values that the two $P$ states are well above the energy of the ground $2^1S$ state (see, e.g., \cite{Sob}). In turn, 
the bound $2^3S$ state of the Be atom ($E$ = -14.430 060 015 $a.u.$ \cite{Fro1}) is located above these two $P$ states. In other words, in the 
four-electron atoms (and ions) the lowest bound state in the singlet series is the ground $2^1S$ state, while the analogous lowest 
bound state in the triplet series is the $2^3P$ state, i.e. the bound state with $L = 1$. 

In this work the total energies of different triplet states in the neutral Be atom have been obtained with the use of the 
Configuration Interaction method (CI) employing $LS$ eigenfunctions and Slater orbitals in the same way utilized in our work of 
Ref. \cite{Li}, except that here we select the $LS$ configurations because their number would be too large. The orbital used 
were $s$, $p$, $d$, $f$, $g$, $h$, and $i$ Slater orbitals defined as $\phi (\mathbf{r}) = r^{n-1}e^{-\alpha r}Y_l^m(\theta ,
\varphi )$. One set of two orbital exponents have been effectively optimized for all configurations. The CI wave function 
takes the form    
\begin{equation}
\Psi =\sum_{p=1}^NC_p\Phi _p,\qquad \Phi _p=\hat{O}(\hat{L}^2)\hat{\mathcal{A%
}} \prod_{k=1}^n\phi _k(r_k,\theta _k,\varphi _k) \chi^{(0)}_1
\end{equation}
$\hat{O}(\hat{L}^2)$ is a projector operator. It indicates that all such configurations are the eigenfunctions of the 
$\hat{L}^2$ operator with the eingenvalue $L (L + 1)$. Among all possible symmetry adapted configurations we have selected the 
ones with an energy contribution to the total energy of the state under consideration is $> 1\cdot 10^{-7} a.u.$, see Table I. In 
the last equation the notation $\chi^{(0)}_1$ is the spin eigenfunction of the triplet state with $S = 1$ and $S_z = 0$, which 
have been chosen in the form  
\begin{equation}
 \chi^{(0)}_1 =\left[ (\alpha \beta -\beta \alpha )(\alpha \beta + \beta \alpha ) \right] \ 
\end{equation}

In this work the total energies of a number of bound triplet $S$, $P$, $D$, $F$, $G$, $H$ and $I$ states have been determined to relatively 
high accuracy. The total energies of these states are shown in Table II. These energies in Table II are ordered with respect to  
their numerical values. As follows from this Table the triplet, rotationally excited states, i.e. $P$, $D$, $F$, $G$, $H$ and $I$ states, are 
slightly more stable than the corresponding $S$ states with the same principal quantum number $n$ ($n$ = 2, 3, 4, 5, $\ldots$). Also, 
the direct comparison of the total energies of the triplet and singlet states would be interesting to study the fulfillment of the 
Hund's rule of maximum multiplicity in the four-electron Be atom. The total energies of some triplet $P$ states in the Be 
atom were evaluated in a few earlier studies using the variational full-core plus correlation wave function (FCPC) \cite{Chung,Chen} 
(note that the non-relativistic variational FCPC energy values are also corrected including the relativistic effects of the $1s^2$ 
electrons, we call this method 'FCPC corrected') and others based on explicitly correlated Monte Carlo (MC) \cite{Galv,Bertini} 
and Multiconfiguration Hartree-Fock methods (MCHF) \cite{Froese}. These calculations are summarized in Table III. Figure 1 shows the 
approximate bound state spectrum of the triplet states of $^9$Be. For this picture we have used our best-to-date results (total 
energies) and best energies known from the literature. The graphical representation of the excited $P$, $D$, $F$, $G$, $H$ and $I$ triplet 
states is approximate and it is based on the CI calculations of this work.    

It is clear that the total energy of an arbitrary bound state in the four-electron Be atom must be lower than the total energy 
of the ground (doublet) $2^2S$ state of the three-electron Be$^{+}$ ion (dissociation or threshold energy $E_{tr}$), which equals 
-14.324 763 176 47 $a.u.$ \cite{Yan}. If (and only if) this condition is obeyed, then such a state is a truly bound state. As follows 
from Table II all states considered in this study are bound. 

\section{General structure of the bound state spectra}

The method described above allows one to conduct accurate computations of different bound states in the four-electron atoms and ions.
Such states include both the singlet and triplet bound states and rotationally excited bound states with arbitrary, in principle, angular
momentum $L$. It opens a new avenue in the study of the bound spectra in four- and many-electron atoms and ions, since in all earlier 
works only one (or very few) bound states with $L \le 1$ were considered. Based on the results of these studies it was very difficult to 
obtain an accurate and realistic picture of the bound state spectra in the four-electron atoms and ions. Now, we can investigate the 
whole bound state spectra of the different four-electron atomic species.

For the triplet states in the Be atom considered in this study we determined the total energies of a large number of bound states. For
better understanding of the relative positions of these bound states we plotted the total energies of these (triplet) bound states in one
diagram (see Fig.1). In old books on atomic spectroscopy such pictures (or diagrams) were called the `spectral diagrams'. Thus, our Fig.1
is the spectral diagram of the triplet states of the ${}^{\infty}$Be atom. Later, we have found that such a spectral diagram  is a very useful 
tool for studing some effects (e.g., electron-electron correlations and electron-electron repulsion) which essentially determine the actual 
order of the bound states in the spectrum and energy differences between them. Formally, by performing numerical calculations of a large number 
of bound states in atomic systems one always needs to answer the two following questions: (1) predict the correct order of low-lying bound 
states, and (2) describe transitions between the low-lying bound states and weakly-bound Rydberg states. To solve the first problem we can 
compare our results with the known experimental data for Be atom \cite{Kramida}. In general, the agreement between our computational results and 
picture (see Fig.1) and data for the beryllium atom presented in \cite{Kramida} can be considered as very good. It is clear we have calculated 
only the non-relativistic (total) energies, i.e. all relativistic and lowest-order QED corrections were ignored. Note also that our CI-method is
substantially more accurate than various procedures based on Hartree-Fock approximation, but it still provides a restricted description of the 
electron-electron correlations in actual atoms and ions. Nevertheless, the observed agreement with the actual bound state spectra of the triplet
states in the Be atom (or ${}^{9}$Be atom) is very good for low-lying bound states. 

The second question formulated above is related to the weakly-bound Rydberg states which were called the `hydrogenic' states in old
atomic physics books. To explain the situation we note that the total energies of such states in the Be atom are described by the following formula
(in atomic units)
\begin{eqnarray}
  E({\rm Be}; n L) = E({\rm Be}^{+}; 2^2S) - \frac{m_e e^4}{2 \hbar^2} \frac{1}{(n + \Delta_{\ell})^2} = 
  -14.324 763 176 47 - \frac{1}{2 (n + \Delta_{\ell})^2} \label{Rydb}
\end{eqnarray}
where $L = \ell$ (in this case), $n$ is the principal quantum number of the $n L$ state ($L$ is the angular quantum number) of the Be atom and 
$\Delta_{\ell}$ is the Rydberg correction which explicitly depends upon $\ell$ (angular momentum of the outer most electron) and the total 
electron spin of this atomic state. It can be shown that the Rydberg correction rapidly vanishes with $\ell$ (for given $n$ and $L$). Moreover, 
this correction $\Delta_{\ell}$ also decreases when the principal quantum number $n$ grows. It follows from this that the energy differences between 
the corresponding singlet and triplet bound states with the same $n$ and $L$ rapidly converges to zero when any of these two quantum numbers 
increase. This criterion is important in applications and allows one to classify all bound states as the Rydberg states, pre-Rydberg and
non-Rydberg states. 

As follows from the results of our calculations all bound states with $n \ge 6$ in the beryllium atom are the weakly-bound, Rydberg states. On the 
other hand, all bound triplet states in the Be atom with $n \ge 4$ can be considered as pre-Rydberg states. To illustrate this we have determined the 
total energies of the triplet $4^{3}F$, $5^{3}G$ states and compared them with the total energies of the singlet $4^{1}F$, $5^{1}G$ states (see Table 
IV). As follows from Table IV the energy difference between the $4^{3}F$ and $4^{1}F$ states is $\approx 9.04 \cdot 10^{-4}$ $a.u.$, while the 
analogous difference between the total energies of the $5^{3}G$ and $5^{1}G$ states is $\approx 1.35 \cdot 10^{-4}$ $a.u.$ The ratios of these 
differences to the threshold energy $E_{tr}$ mentioned above are $\approx 6.31 \cdot 10^{-5}$ and $9.42 \cdot 10^{-6}$, respectively. 
These numerical values are very small in comparison with the unity. On the other hand, these values are larger than the value $1 \cdot 10^{-5}$ 
which can be found for the actual Rydberg states. Therefore, the bound states of the Be atom with $n = 4$ and $n = 5$ can be considered as the 
pre-Rydberg states.  
   
\section{Variational expansion in multi-dimensional gaussoids}

It should be mentioned here that some bound $n^3S$ and $n^3P$ states in the Be atom have also been evaluated (see below) with the use of the basis 
set of multi-dimensional gaussoids proposed by Kolesnikov and Tarasov in the middle of 1970's for nuclear few-body problems. We shall call  
this wave function in the next section the 'KT-expansion' \cite{KT}. This method is different from the CI method mentioned above, 
but it allows one to evaluate very accurately total energies for some low-lying $S$ and $P$ states in the four-electron atoms and ions. 
This method is discussed below for the bound triplet ${}^3S$ states in the Be atom. 

First, note that the wave function of an arbitrary bound ${}^3S$ states in the Be atom can always be represented as the sum of products of the radial 
and spin functions (or configurations) \cite{LLQ}. For a triplet we use the spin eigenfunction with $S = 1$ and $S_z = 1$, where $S$ is the total electron 
spin, i.e. ${\bf S} = {\bf s}_1 + {\bf s}_2 + {\bf s}_3 + {\bf s}_4$, of four-electrons and $S_z$ is its $z-$projection. Therefore, our spin function 
$\chi_{11}(1,2,3,4)$ is defined by the following equalities: ${\bf S}^2 \chi_{11}(1,2,3,4) = 1 (1 + 1) \chi_{11}(1,2,3,4) = 2 \chi_{11}(1,2,3,4)$ and $S_z 
\chi_{11}(1,2,3,4) = \chi_{11}(1,2,3,4)$. In general, there are two such spin functions for each four-electron atom/ion in the triplet state. Below, we 
chose such functions in the form $\chi^{(1)}_{11} = (\alpha \beta - \beta \alpha) \alpha \alpha$ and $\chi^{(2)}_{11} = (2 \alpha \alpha \beta - \beta 
\alpha \alpha - \alpha \beta \alpha) \alpha$, where the notations $\alpha$ and $\beta$ denote spin-up and spin-down functions \cite{LLQ}, respectively. 
The total four-electron wave function of the triplet states is represented as the following sum  
\begin{eqnarray}
 \Psi = {\cal A}_e [\psi(A;\{r_{ij}\}) (\alpha \beta - \beta \alpha) \alpha \alpha] + {\cal A}_e [\phi(B;\{r_{ij}\}) (2 \alpha 
 \alpha \beta - \beta \alpha \alpha - \alpha \beta \alpha) \alpha] \label{eq2}
\end{eqnarray}
where the notation $\{r_{ij}\}$ designates all ten relative coordinates (electron-nuclear and electron-electron coordinates) in the 
four-electron Be atom, while the notation ${\cal A}_e$ means the complete four-electron antisymmetrizer. The explicit formula for 
the ${\cal A}_e$ operator is 
\begin{eqnarray}
 {\cal A}_e = \hat{e} - \hat{P}_{12} - \hat{P}_{13} - \hat{P}_{23} - \hat{P}_{14}
 - \hat{P}_{24} - \hat{P}_{34} + \hat{P}_{123} + \hat{P}_{132} + \hat{P}_{124} 
 + \hat{P}_{142} + \hat{P}_{134} + \hat{P}_{143} \nonumber \\ 
 + \hat{P}_{234} + \hat{P}_{243} 
 - \hat{P}_{1234} - \hat{P}_{1243} - \hat{P}_{1324} - \hat{P}_{1342} - \hat{P}_{1423} 
 - \hat{P}_{1432} + \hat{P}_{12} \hat{P}_{34} + \hat{P}_{13} \hat{P}_{24} 
 + \hat{P}_{14} \hat{P}_{23} \label{eq3}
\end{eqnarray}
Here $\hat{e}$ is the identity permutation, while $\hat{P}_{ij}$ is the permutation of the spin and spatial coordinates of the 
$i$th and $j$th identical particles. Analogously, the notations $\hat{P}_{ijk}$ and $\hat{P}_{ijkl}$ stand for the consecutive 
permutations of the spin and spatial coordinates of the three and four identical particles (electrons). In real bound state 
calculations one needs to know the explicit expressions for the spatial projectors only. 

These spatial projectors are usualy obtained by applying the ${\cal A}_e$ operator to each component of the wave function in 
Eq.(\ref{eq2}). At the second step we need to determine the scalar product (or spin integral) of the result and incident spin 
function. After the integration over all spin variables one finds the corresponding spatial projector. For instance, in the case 
of the first term in Eq.(\ref{eq2}) we obtain
\begin{eqnarray}
 {\cal P}_{\psi\psi} = \frac{1}{2 \sqrt{6}} (2 \hat{e} + 2 \hat{P}_{12} - \hat{P}_{13} - 
 \hat{P}_{23} - \hat{P}_{14} - \hat{P}_{24} - 2 \hat{P}_{34} - 2 \hat{P}_{12} \hat{P}_{34}
 - \hat{P}_{123} - \hat{P}_{124} - \hat{P}_{132} \nonumber \\
 - \hat{P}_{142} + \hat{P}_{134} + \hat{P}_{143} + \hat{P}_{234} + \hat{P}_{243} 
 + \hat{P}_{1234} + \hat{P}_{1243} + \hat{P}_{1342} + \hat{P}_{1432}) \label{eq4}
\end{eqnarray}
Analogous formulas have been found for two other spatial projectors ${\cal P}_{\psi\phi} = {\cal P}_{\phi\psi}$ and ${\cal P}_{\phi\phi}$. 
The explicit formulas for the ${\cal P}_{\psi\phi} = {\cal P}_{\phi\psi}$ and ${\cal P}_{\phi\phi}$ spatial projectors are significantly 
more complicated and not presented here. These formulas can be requested from the authors. In actual bound state calculations we can 
always restrict ourselves to one spin function, i.e. $\chi^{(1)}_{11}$, (or one spin configuration) and use the formula, Eq.(\ref{eq4}). 

The functions $\psi(A;\{r_{ij}\})$ and $\phi(B;\{r_{ij}\})$ in Eq.(\ref{eq2}) are the radial parts (or components) of the total wave function 
$\Psi$. For the bound states in various five-body systems these functions are approximated with the use of the KT-variational expansion 
written in ten-dimensional gaussoids \cite{KT}, e.g., for the $\psi(A;\{r_{ij}\})$ function we have
\begin{eqnarray}
 \psi(A;\{r_{ij}\}) = {\cal P} \sum^{N_A}_{k=1} C_K \exp(-\sum_{ij} a_{ij} r^{2}_{ij}) 
 \label{eq5}
\end{eqnarray}
where $N_A$ is the total number of terms, $C_k$ are the linear variational coefficients and ${\cal P}$ is the spatial projector defined in 
Eq.(\ref{eq4}). The notations $A$ (or notations $A$ and $B$ in Eq.(\ref{eq2})) stands for the corresponding set of the non-linear parameters 
$\{ a^{(k)}_{ij} \}$ (and $\{ b^{(k)}_{ij} \}$) in the radial wave functions, Eq.(\ref{eq5}) (or Eq.(\ref{eq2})). It is assumed here that these 
two sets of non-linear parameters $A$ and $B$ are optimized independently of each other. 

In general, the KT-variational expansion is very effective for various few-body systems known in atomic, molecular and nuclear physics. A large 
number of fast algorithms developed recently for optimization of the non-linear parameters in the trial wave functions, Eq.(\ref{eq5}), allow one 
to approximate the total energies and variational wave functions to high accuracy. In some cases, however, the overall convergence rate for some 
bound state properties is considerably lower than for the total energies and other properties. In particular, it was found that the expectation 
values of the two- and three-particle delta-functions, e.g., $\langle \delta({\bf r}_{eN}) \rangle$, converge slowly. In other words, it takes a 
long time to approximate these expectation values to high accuracy. The total energy and other expectation values do not change drastically during 
such an additional optimization of the non-linear parameters in the wave functions. In this study we report a number of expectation values computed 
for the bound triplet $2^3S, 3^3S, 4^3S$ and $5^3S$ states of the Be atom (see Table IV) computed with the use of KT-expansion. The expectation 
values of the electron-nuclear delta-function $\langle \delta({\bf r}_{eN}) \rangle$ for these states are important to perform approximate numerical 
evaluation of the hyperfine structure splitting for each of these states in the ${}^{9}$Be atom. 

\section{Bound state properties}

Variational KT-expansion in multi-dimensional gaussoids allows one to obtain very accurate numerical values of the total energies. The fastest  
convergence of the KT-expansion is observed for the low-lying ${}^{3}S$ states. In this study by using the KT-expansion we have determined the total 
energies and some other bound state properties of a few low-lying triplet ${}^{3}S$ states in the ${}^{\infty}$Be atom. The current total energies 
of these states are: $E$ = -14.430 060 025 $a.u.$ ($2^3S$ state), -14.372 858 590 $a.u.$ ($3^3S$ state), -14.351 112 516 $a.u.$ ($4^3S$ state), 
-14.337 598 153 $a.u.$ ($5^3S$ state) and -14.328 903 15 $a.u.$ ($6^3S$ state). In general, the total energies of the $n^{3}S-$states are less 
accurate than our total energy obtained for the $2^3S$ state. 

By using our variational wave functions constructed in the KT-method we can determine a number of bound state properties of the Be atom in these 
bound states. The bound state properties of the $2^3S$, $3^3S, 4^3S$ and $5^3S$ states can be found in Tables IV and V. The expectation values from 
Tables IV and V represent the basic geometrical and dynamical propeties of the four-electron Be atom in these $S$ states. The physical meaning of all 
notations used in Tables IV and V to designate the bound state properties (or expectation values) is clear. All expectation values in these Tables are 
given in atomic units ($\hbar = 1, m_e = 1$ and $e = 1$). Note that these expectation values have never been determined in earlier studies. Moreover,
these expectation values can be used in various applications, e.g., electron-nuclear delta-functions are needed to determine the hyperfine structure 
splittings in the triplet $S$ states (see, e.g., \cite{Fro1}). Another interesting problem is to study changes in the bound state properties of the 
triplet $S$ states which occur when the electric charge of the central nucleus $Q$ increases. Since the parameter $Q$ in Eq.(\ref{eq1}) is an integer 
number, then we can write for an arbitrary expectation value $\langle X \rangle$ 
\begin{eqnarray}
 \langle X(Q) \rangle = a_2 Q^2 + a_1 Q + a_0 + b_1 Q^{-1} +  b_2 Q^{-2} + b_3 Q^{-3} +  b_4 Q^{-4} + \ldots \label{Loran}
\end{eqnarray}
where the coefficients $a_2, a_1, a_0$ and $b_1, b_2, b_3, b_4, \ldots$ are the real numbers and $Q \ge 4$ for the bound triplet states in the 
four-electron atoms/ions. The first three terms in Eq.(\ref{Loran}) form the regular part of this expansion (series), while all terms with the 
coefficients $b_{i}$ ($i = 1, 2, \ldots$) form the principal part of the expansion, Eq.(\ref{Loran}). The explicit form of expansion Eq.(\ref{Loran}) 
follows (see, e.g., \cite{Epst}) from the fact that the non-relativistic Coulomb Hamiltonian Eq.(\ref{eq1}) is a quadratic form of the electron momenta 
${\bf p}_i$ ($i = 1, 2, \ldots, N$) for $N-$electron atoms. In actual applications to atomic systems the formula, Eq.(\ref{Loran}), is used to predict the 
bound state properties, including the total energies, for those atomic systems for which direct numerical calculations is either difficult, or impossible. In 
reality, the numerical values of the coefficients $a_2 ,a_1, a_0$ and $b_1, b_2, b_3, \ldots$ are determined by fitting the results of accurate numerical 
computations with Eq.(\ref{Loran}). 

In this study by using the total energies of a number of bound $2^3S-$states in the four-electron Be atom and some Be-like ions (B$^{+}$, C$^{2+}$, \ldots, 
Na$^{7+}$ and Mg$^{8+}$) we determine the coefficients $a_2 ,a_1, a_0$ and $b_1, b_2, b_3$ for the total energies. In other words, we determine the actual 
six-term expansion, Eq.(\ref{Loran}), for the function $E(Q)$ (see Table VI). Note that the total energies of the $2^3S$ states in these ions (see Table VI) 
are now known to much better accuracy, than it followed from earlier studies. Therefore, we can make a conclusion that our coefficients $a_2 ,a_1, a_0, b_1, 
b_2$ and $b_3$ are also substantially more accurate and reliable than values obtained in earlier works. In principle, by using the same formula, Eq.(\ref{Loran}), 
we can evaluate other bound state properties for various bound states in the four-electron ions.  

\section{Conclusion}

We have investigated the bound state spectrum of the low-lying triplet states in the four-electron Be atom. The total energies of the bound $S(L = 0)$, $P(L = 
1)$, $D(L = 2)$, $F(L = 3)$, $G(L = 4)$, $H(L = 5)$ and $I(L = 6)$ states have been determined to the accuracy $\approx 1 \cdot 10^{-3} - 2 \cdot 10^{-3}$ a.u. which is 
much better than methods based on the Hartree-Fock approximation can provide. The results of our calculations are accurate and they agree with the experimental data known 
for these states from \cite{Kramida}. Note that our advanced approach has no restrictions in applications and allows one to investigate the whole spectrum of bound 
(triplet) states in the four-electron Be atom and Be-like ions. Such an analysis includes rotationally excited and highly excited states, and states with the different 
(total) electron spin. By using our method it is possible to observe and investigate the actual transition from the low-lying bound bound states to the weakly-bound (or 
Rydberg) states in the spectra of the four-electron Be atom. These important advantages of our approach allowed us to draw the first spectral diagram of the triplet 
states in the four-electron beryllium atom. It appears that such spectral diagrams can be very useful in theoretical research and experimental applications. 

Briefly, we can say that this work opens a new avenue in accurate numerical analysis of the bound triplet states in four-electron atoms and ions. It is expected that 
other theoretical and experimental works on the triplet states in four-electron atoms and ions will follow. In particular, in our next study we want to consider the 
low-lying triplet bound states in the four-electron, Be-isoelectronic ions, including such ions of boron, carbon and nitrogen which are important in stellar astrophysics.
An obvious achievement of our study for the few-body physics is the analysis of the whole spectra of bound (triplet) states in the four-electron atom(s), while in all
earlier works only a very few bound states were investigated.

\newpage


\begin{center}
  \sansmath
  \begin{tikzpicture}[
    font=\sffamily,
    level/.style={black,thick},
    ionization/.style={black,dashed},
  ]
  \coordinate (sublevel) at (0, 8pt);

  \node at (0.5, 10.5) {$^3$S};
  \node at (2.5, 10.5) {$^3$P};
  \node at (4.5, 10.5) {$^3$D};
  \node at (6.5, 10.5) {$^3$F};
  \node at (8.5, 10.5) {$^3$G};
  \node at (10.5, 10.5) {$^3$H};
  \node at (12.5, 10.5) {$^3$I};
  \node at (-0.25,10.0) {n};

  \node at (0.5, 5.46) {\scriptsize $2s3s$};
  \node at (0.5, 7.40) {\scriptsize $2s4s$};
  \node at (0.5, 8.70) {\scriptsize $2s5s$};
  \node at (0.5, 9.50) {\scriptsize $2s6s$};
  \coordinate (S00) at (0, 5.66);
  \coordinate (S10) at (0, 7.6);
  \coordinate (S20) at (0, 8.90);
  \coordinate (S30) at (0, 9.70);

   \foreach \level/\text in { 00/2,  10/3,  20/4,  30/5}
     \draw[level] (S\level) node[left=20pt] {} node[left]
        {\footnotesize {\text}} -- +(1.0, 0);

  \node at (2.5,-0.20) {\scriptsize $2s2p$};
  \node at (2.5, 6.74) {\scriptsize $2s3p$};
  \node at (2.5, 8.25) {\scriptsize $2s4p$};
  \node at (2.5, 9.30) {\scriptsize $2s5p$};
  \coordinate (P00) at (2, 0.0);
  \coordinate (P10) at (2, 6.94);
  \coordinate (P20) at (2, 8.45);
  \coordinate (P30) at (2, 9.5);

  \foreach \level/\text in {00/2, 10/3, 20/4, 30/5}
    \draw[level] (P\level) node[left=20pt] {} node[left]
     {\footnotesize {\text}} -- +(1.0, 0);

  \node at (4.5,7.36) {\scriptsize $2s3d$};
  \node at (4.5,8.40) {\scriptsize $2s4d$};
  \node at (4.5,9.35) {\scriptsize $2s5d$};
  \coordinate (D00) at (4, 7.56);
  \coordinate (D10) at (4, 8.6);
  \coordinate (D20) at (4, 9.55);

  \foreach \level/\text in {00/3, 10/4, 20/5}
    \draw[level] (D\level) node[left=20pt] {} node[left]
     {\footnotesize {\text}} -- +(1.0, 0);

  \node at (6.5,8.60) {\scriptsize $2s4f$};
  \node at (6.5,9.40) {\scriptsize $2s5f$};
  \coordinate (F00) at (6, 8.80);
  \coordinate (F10) at (6, 9.60);

  \foreach \level/\text in {00/4, 10/5 }
    \draw[level] (F\level) node[left=20pt] {} node[left]
     {\footnotesize {\text}} -- +(1.0, 0);

  \node at (8.5,9.45) {\scriptsize $2s5g$};
  \coordinate (G00) at (8,9.65);

  \foreach \level/\text in {00/5}
    \draw[level] (G\level) node[left=20pt] {} node[left]
     {\footnotesize {\text}} -- +(1.0, 0);

  \draw[ionization] (0, 10.0) node[left=20pt] {E(${\rm Be}^+$)}-- +( 13, 0);

  \draw[level] (0,9.8) node[left=20pt] {}-- +(1.0, 0);
  \draw[level] (0,9.9) node[left=20pt] {}-- +(1.0, 0);

  \draw[level] (2,9.8) node[left=20pt] {}-- +(1.0, 0);
  \draw[level] (2,9.9) node[left=20pt] {}-- +(1.0, 0);

  \draw[level] (4,9.8) node[left=20pt] {}-- +(1.0, 0);
  \draw[level] (4,9.9) node[left=20pt] {}-- +(1.0, 0);

  \draw[level] (6,9.8) node[left=20pt] {}-- +(1.0, 0);
  \draw[level] (6,9.9) node[left=20pt] {}-- +(1.0, 0);

  \draw[level] (8,9.8) node[left=20pt] {}-- +(1.0, 0);
  \draw[level] (8,9.9) node[left=20pt] {}-- +(1.0, 0);

  \draw[level] (10,9.8) node[left=20pt] {}-- +(1.0, 0);
  \draw[level] (10,9.9) node[left=20pt] {}-- +(1.0, 0);

  \draw[level] (12,9.9) node[left=20pt] {}-- +(1.0, 0);

  \node at (10.5,9.60) {\scriptsize $2s6h$};
  \node at (12.5,9.70) {\scriptsize $2s7i$};

  \end{tikzpicture}
  \captionof{figure}{The energy levels of the triplet states in the beryllium atom. The threshold
energy (or ionization limit) $E($Be$^{+})$ coincides with the total energy of the
ground 2$^2S$ state of the three-electron Be$^{+}$ ion.}
\end{center}

\newpage


\begin{table}[tp]
\begin{center}
\caption{List of of the different $L$ configurations employed in the CI calculations of the $S$, $P$, $D$, $F$, $G$, and 
$I$ states. In all configurations $M=0$. Here it is taken into account that the exponents within a shell have been kept equal.}
\begin{tabular}{|c | c | c | c |}
\hline\hline
State & $L$ & $M$ & Configurations \\
\hline
$S$  & 0 & 0 & $ssss$, $ppss$, $ddss$, $ffss$, $sspp$, $spps$, $pppp$, $ssdd$, $sdds$, $sppd$, $dpps$, \\
     &   &   & $sdpp$, $ppds$, $ppdd$, $dddd$, $ddpp$, $ssff$, $sffs$, $ppff$, $ddff$, $ffpp$, $ffdd$  \\
\hline
$P$  & 1 & 0 & $sssp$, $spss$, $ppsp$, $ddsp$, $ffsp$, $sppp$, $sddp$, $sffp$, $sspd$, $spds$, $sdps$, $pdss$, \\
     &   &   & $pppd$, $ddpd$, $ffpd$, $dfss$, $ssdf$, $sdfs$, $pdpp$, $pddd$, $pdff$, $ppdf$, $dddf$, $ffdf$  \\
\hline
$D$  & 2 & 0 & $sssd$, $sdss$, $sspp$, $spps$, $pppp$, $ppsd$, $ssdd$, $ddsd$, $ffsd$, \\
     &   &   & $ssff$, $sppd$, $sddd$, $sffd$, $sdds$, $ddss$, $sffs$, $ffss$ \\
\hline
$F$  & 3 & 0 & $sssf$, $sfss$, $ppsf$, $ddsf$, $ffsf$, $sppf$, $sddf$ \\
\hline
$G$  & 4 & 0 & $sssg$, $sgss$, $ppsg$, $ddsg$, $ffsg$, $ggsg$, $sppg$, $sddg$ \\
\hline
$H$  & 5 & 0 & $sssh$, $ppsh$, $ddsh$, $ffsh$, $ggsh$, $hhsh$, $spph$, $sddh$ \\
\hline
$I$  & 6 & 0 & $sssi$, $ppsi$, $ddsi$, $ffsi$, $ggsi$, $hhsi$, $iisi$, $sppi$, $sddi$ \\
\hline\hline
\end{tabular}
\end{center}
\end{table}


\begin{table}[tp]
\begin{center}
\caption{The total energies in $a.u.$ of some low-lying triplet states of the ${}^{\infty}$Be atom determined
         with the use of the CI method using $\approx 2000$ configurations. 
         The total energies of the bound states in the upper part of this Table
         are lower than the ionization threshold of the ${}^{\infty}$Be atom ($E_{{}^{\infty}{\rm Be}^{+}}$ =
         -14.324 763 176 47 $a.u.$).}
\begin{tabular}{| c  | c  |  c  |  c  |}
\hline\hline
 State &  $E$   \\
\hline
2$^3P$ & -14.565 365 \\
2$^3S$ & -14.428 858 \\
\hline
3$^3P$ & -14.395 471 \\
3$^3D$ & -14.381 020 \\
3$^3S$ & -14.371 277 \\
\hline
4$^3P$ & -14.357 020 \\
4$^3D$ & -14.350 318 \\
4$^3F$ & -14.353 111 \\
4$^3S$ & -14.345 918 \\
\hline
5$^3P$ & -14.333 510 \\      
5$^3D$ & -14.343 621 \\
5$^3F$ & -14.334 658 \\
5$^3G$ & -14.335 897 \\
5$^3S$ & -14.330 909 \\
\hline
6$^3H$ & -14.327 136 \\
7$^3I$ & -14.322 647 \\
\hline
4$^1F$ & -14.352 207 \\
5$^1G$ & -14.335 762 \\
\hline
$E_{{}^{\infty}{\rm Be}^{+}}$ & -14.324 763 176 \\
\hline\hline
\end{tabular}
\end{center}
\end{table}


%
\begin{table}[tp]
\begin{center}
\caption{Comparison with other calculations using the full-core plus correlation method (FCPC), 
variational Monte Carlo (VMC) and diffusion Monte Carlo (DMC). In the corrected FCPC relativistic 
corrections have been added. All energies are given in $a.u.$}  
\begin{tabular}{|c |l| l| c|}                                 
\hline\hline
State  &  Method               &     $E$        & Ref.         \\                                        
\hline
2$^3P$ & FCPC  $\psi_0$        & -14.562 231 00   & \cite{Chung} \\                 
       & VMC                   & -14.564 65(3)    & \cite{Galv}  \\
       & CI(STO)               & -14.565 365      & This work \\
       & FCPC                  & -14.566 37       & \cite{Chen} \\   
       & MCHF                  & -14.566 560      & \cite{Froese} \\                          
       & FCPC $\psi_0+\psi_1$  & -14.566 913 79   & \cite{Chung} \\
       & FCPC corrected        & -14.567 238 30   & \cite{Chung} \\
\hline 
2$^3S$ & FCPC  $\psi_0$        & -14.426 290 63   & \cite{Chung} \\                  
       & VMC                   & -14.427 46(4)    & \cite{Galv}  \\        
       & CI(STO)               & -14.428 858      & This work    \\
       & FCPC, $\psi_0+\psi_1$ & -14.429 798 54   & \cite{Chung} \\ 
       & multi-dimensional Gaussoids & -14.430 060 025 & This work  \\                                             
       & FCPC corrected        & -14.430 064 40   & \cite{Chung} \\                                               
\hline
3$^3P$ & FCPC  $\psi_0$        & -14.395 099 06   & \cite{Chung} \\                
       & MC                    & -14.395 31(4)    & \cite{Galv}  \\
       & DMC                   & -14.395 47(5)    & \cite{Bertini} \\
       & CI(STO)               & -14.395 471      & This work \\
       & FCPC                  & -14.398 39       & \cite{Chen} \\                                                     
       & FCPC $\psi_0+\psi_1$  & -14.398 646 79   & \cite{Chung} \\                
       & FCPC corrected        & -14.398 960 12   & \cite{Chung} \\                                          
\hline
3$^3D$ & FCPC  $\psi_0$        & -14.379 559 32   & \cite{Chung} \\      
       & CI(STO)               & -14.381 020      & This work \\ 
       & MC                    & -14.382 19(3)    & \cite{Galv}  \\    
       & FCPC $\psi_0+\psi_1$  & -14.384 356 30   & \cite{Chung} \\               
       & FCPC corrected        & -14.384 622 60   & \cite{Chung} \\ 
\hline
3$^3S$ & CI(STO)                     & -14.371 277     &This work \\ 
       & multi-dimensional Gaussoids & -14.372 858 590 & This work \\
\hline\hline
\end{tabular}
\end{center}
\end{table}

\begin{table}[tp]
\begin{center}
{Table III. Continuation.}\\ 
\begin{tabular}{|c| l| l| c|}
\hline         
State  &  Method                     &     $E$        & Ref.         \\
\hline
\hline
4$^3P$ & CI(STO)                     & -14.357 020      & This work \\
       & FCPC                        & -14.362 48       & \cite{Chen} \\
\hline
4$^3D$ & CI(STO)                     & -14.350 318      & This work   \\
4$^1F$ & CI(STO)                     & -14.352 207      & This work   \\
4$^3F$ & CI(STO)                     & -14.353 111      & This work   \\
4$^3S$ & CI(STO)                     & -14.345 918      & This work \\ 
4$^3S$ & multi-dimensional Gaussoids & -14.351 112 52   & This work \\
5$^3S$ & multi-dimensional Gaussoids & -14.337 598 15   & This work \\
5$^1G$ & CI(STO)                     & -14.335 762      & This work    \\
5$^3G$ & CI(STO)                     & -14.335 897      & This work    \\
6$^3S$ & multi-dimensional Gaussoids & -14.328 903 15    & This work \\
\hline\hline
\end{tabular}
\end{center}
\end{table}


 \begin{table}[tbp]
   \caption{The expectation values of some electron-nuclear ($en$) and electron-electron ($ee$) 
            properties of the $2^3S$ state of the ${}^{\infty}$Be atom (in atomic units).
            The notation $N$ stands for the total number of basis functions used in calculations.}
     \begin{center}
     \begin{tabular}{| c | c | c | c | c | c | c |}
       \hline\hline          
 $N$ & $\langle r^{-2}_{eN} \rangle$ & $\langle r^{-1}_{eN} \rangle$ & $\langle r_{eN} \rangle$ & $\langle r^2_{eN} \rangle$  &  $\langle r^3_{eN} \rangle$  &  $\langle r^4_{eN} \rangle$ \\
     \hline
 600 & 14.2748 593 & 2.0359 911 & 2.6329 155 & 17.183 562 & 148.4227 & 1470.159 \\

 800 & 14.2748 644 & 2.0359 907 & 2.6329 097 & 17.183 578 & 148.4245 & 1470.200 \\
      \hline          
 $N$ & $\langle r^{-2}_{ee} \rangle$ & $\langle r^{-1}_{ee} \rangle$ & $\langle r_{ee} \rangle$ & $\langle r^2_{ee} \rangle$ &  $\langle r^3_{ee} \rangle$ &  $\langle r^4_{ee} \rangle$ \\
     \hline
 600 & 1.5034 356 & 0.6192 659 & 2.6329 155 & 35.207 06 & 320.2519 & 3281.012 \\

 800 & 1.5034 362 & 0.6192 667 & 2.6329 097 & 35.207 12 & 320.2562 & 3281.110 \\
     \hline              
 $N$ &  $\langle \frac12 p^2_{e} \rangle$ & $\langle \frac12 p^2_{N} \rangle$ & $\langle \delta_{eN} \rangle$ & $\langle \delta_{ee} \rangle$ & $\langle \delta_{eeN} \rangle$ & $\langle \delta_{eeeN} \rangle$ \\ 
     \hline
 600 & 3.6075 575 & 14.892 986 & 8.739 445 & 0.2650 593 & 34.93691 & 0.0 \\

 800 & 3.6075 586 & 14.892 992 & 8.739 451 & 0.2650 594 & 34.93697 & 0.0 \\
     \hline\hline
  \end{tabular}
  \end{center}
  \end{table}


 \begin{table}[tbp]
   \caption{The expectation values of some electron-nuclear ($en$) and electron-electron ($ee$) 
            properties of the $3^3S$, $4^3S$ and $5^3S$ states of the ${}^{\infty}$Be atom (in atomic units).
            The notation $N$ stands for the total number of basis functions used in calculations.}
     \begin{center}
     \begin{tabular}{| c | c | c | c | c | c | c |}
       \hline\hline          
 state & $\langle r^{-2}_{eN} \rangle$ & $\langle r^{-1}_{eN} \rangle$ & $\langle r_{eN} \rangle$ & $\langle r^2_{eN} \rangle$  &  $\langle r^3_{eN} \rangle$  &  $\langle r^4_{eN} \rangle$ \\
     \hline
 $3^3S$ & 14.2581 303 & 2.0154 403 & 4.6448 319 & 69.099 07 & 1275.485 & 25527.53 \\

 $4^3S$ & 14.2557 019 & 2.0089 059 & 7.1634 186 & 188.09 14 & 5870.504 & 195381.5 \\

 $5^3S$ & 14.2582 045 & 2.0059 496 & 10.068 762 & 401.66 96 & 18577.67 & 908943.3 \\
      \hline          
 state & $\langle r^{-2}_{ee} \rangle$ & $\langle r^{-1}_{ee} \rangle$ & $\langle r_{ee} \rangle$ & $\langle r^2_{ee} \rangle$ &  $\langle r^3_{ee} \rangle$ &  $\langle r^4_{ee} \rangle$ \\
     \hline
 $3^3S$ & 1.4910 864 & 0.582 333 & 8.6633 016 & 138.467 46 & 2586.470 & 52122.14 \\

 $4^3S$ & 1.4900 693 & 0.569 099 & 13.684 320 & 376.301 84 & 11794.71 & 393453.4 \\

 $5^3S$ & 1.4914 313 & 0.562 578 & 19.488 137 & 803.383 46 & 37229.76 & 1823356. \\
     \hline              
 state &  $\langle \frac12 p^2_{e} \rangle$ & $\langle \frac12 p^2_{N} \rangle$ & $\langle \delta_{eN} \rangle$ & $\langle \delta_{ee} \rangle$ & $\langle \delta_{eeN} \rangle$ & $\langle \delta_{eeeN} \rangle$ \\ 
     \hline
 $3^3S$ & 3.5951 891 & 14.8334 114 & 8.662 626 & 0.265 321 & 36.8164 & 0.0 \\ 

 $4^3S$ & 3.5946 643 & 14.8199 863 & 8.578 452 & 0.265 989 & 36.3418 & 0.0 \\

 $5^3S$ & 3.5962 032 & 14.8174 746 & 8.558 327 & 0.267 768 & 35.9772 & 0.0 \\
     \hline\hline
  \end{tabular}
  \end{center}
  \end{table}


 \begin{table}[tbp]
   \caption{The total energies $E$ of the $2^3S$ states in a number of four-electron (Be-like) ions (in $a.u.$) and coefficients of the $Q^{-1}$ 
            expansion constructed for the total energies of the $2^{3}S$ states in these four-electron ions. Expansions with $N$ = 5 and 6 terms 
            are considered.}
     \begin{center}
     \begin{tabular}{| c | c | c | c | c |}
       \hline\hline          
 Ion & $Q$  & $E$ & $C_{i}^{(a)} (N = 5)$ & $C_{i}^{(a)} (N = 6)$ \\
     \hline
  Be        & 4  & -14.430 060 025 & -1.180 657 409 1 & -1.180 491 428 7 \\

  B$^{+}$   & 5  & -23.757 501 147 &  1.293 393 850 7 &  1.287 374 106 6 \\

  C$^{2+}$  & 6  & -35.449 661 045 & -0.717 987 039 3 & -0.633 773 069 6 \\

  N$^{3+}$  & 7  & -49.504 378 033 &  0.170 745 208 9 & -0.397 200 043 1 \\

  O$^{4+}$  & 6  & -65.920 988 652 & -0.605 053 278 7 &  1.242 477 680 8 \\

  F$^{5+}$  & 7  & -84.699 246 319 &               & -2.321 620 789 9 \\

  Ne$^{6+}$ & 8  & -105.838 762 53 &               & \\

  Na$^{7+}$ & 9  & -129.339 671 77 &               & \\

  Mg$^{8+}$ & 10 & -155.201 891 91 &               & \\
     \hline\hline
  \end{tabular}
  \end{center}
${}^{(a)}$Here $C_1 =  a_2,  C_2 =  a_1, C_3 =  a_0, c_4 = b_1, c_5 = b_2, \ldots$ in Eq.(\ref{Loran}).
  \end{table}

\end{document}